\documentclass[aps,twocolumn,showpacs,amsmath,amssymb]{revtex4}
\usepackage{graphicx}
\usepackage{dcolumn}
\usepackage{bm}
\begin{document}

\title{On the fate of travelling waves at the boundary of quantum droplets}
\author{Angel Paredes$^1$, Jose Guerra-Carmenate$^{2,1}$,  and Humberto Michinel$^1$}
\affiliation{$^1$Instituto de F\'\i sica e Ciencias Aeroespaciais (IFCAE), 
Universidade de Vigo, 32004 Ourense, Espa\~na.\\
$^2$Instituto Universitario de Matem\'atica Pura y Aplicada (IUMPA),
Universitat Polit\`ecnica de Val\`encia. Edificio 8E, 46022 Valencia, Spain.\\
}

\begin{abstract}
We analyze quantum droplets formed in a two-dimensional symmetric mixture of Bose-Einstein condensed atoms. For sufficiently large atom numbers, these droplets exhibit a flat-top density profile with sharp boundaries governed by surface tension. Within the bulk of the droplet, traveling matter waves---localized density dips---can propagate at constant velocity while maintaining their shape. Using numerical simulations and qualitative analysis, we investigate the rich phenomenology that arises when such excitations reach the boundary of a finite droplet. We show that they can emit a small outgoing droplet, excite internal modes of the host soliton, or, in the case of vortex-antivortex pairs, split into individual vortices propagating backward near the edge. Furthermore, we demonstrate that traveling waves can be dynamically generated near the boundary through the collision of distinct droplets, and we discuss their trajectories and interactions.

\end{abstract}
\maketitle

\section{Introduction}

The concept of quantum droplets (QDs) in dilute ultracold atomic gases emerged shortly after the realization of Bose-Einstein condensation (BEC). In a seminal proposal \cite{bulgac2002dilute}, it was hypothesized that a BEC could become self-bound --- i.e., stable without an external trapping potential --- if interatomic interactions were finely tuned to balance the competing effects of dispersion and attraction. This would give rise to stable, isolated, shape-preserving atomic clouds with novel physical properties.

However, such self-bound states remained elusive for many years. While BECs are inherently quantum systems, their dynamics are often well described by the mean-field Gross-Pitaevskii equation (GPE) \cite{dalfovo1999theory}. This equation, with attractive interactions, admits localized solutions in one dimension, but such solitonic states are unstable in higher dimensions and prone to collapse \cite{sulem2007nonlinear,kartashov2019frontiers,malomed2024multidimensional}.

A major breakthrough came in 2015, when Petrov \cite{petrov2015quantum} proposed a stabilization mechanism based on the Lee-Huang-Yang (LHY) correction \cite{lee1957}, which incorporates quantum fluctuations into the mean-field framework. This additional repulsive term can counteract collapse in three dimensions, enabling the formation of stable QDs under suitable conditions. Following this proposal, several experimental groups rapidly confirmed the existence of stable atomic droplets in various setups, including dipolar gases \cite{kadau2016observing,ferrier2016observation,schmitt2016self,chomaz2016quantum}, homonuclear mixtures \cite{cheiney2018bright,cabrera2018,semeghini}, and heteronuclear BEC mixtures \cite{d2019observation,burchianti2020dual,politi2022interspecies}.

These advances have sparked intense theoretical activity focused on the modified GPE with the LHY correction, aimed at understanding both the static and dynamical properties of QDs. This equation defines a nonlinear wave system with a rich and complex dynamical landscape; see \cite{bottcher2020new,luo2021new,guo2021new} for reviews on the subject from various perspectives.

The present work contributes to this ongoing effort by addressing a specific question in a simplified setting. We focus on the two-dimensional, symmetric case first discussed in \cite{petrov2016ultradilute}. The effective dimensional reduction from three to two dimensions arises from assuming tight confinement along the $z$-axis, yielding a pancake-shaped atomic cloud (the crossover between the 3D and 2D cases was studied in detail in \cite{pelayo2025phases}). The symmetric configuration corresponds to a homonuclear BEC mixture with equal populations and symmetric initial conditions. Under these assumptions, the coupled system can be effectively reduced to a single partial differential equation. This theoretical model, presented in Section~\ref{sec:math}, provides one of the simplest frameworks for analyzing key properties of QDs, and the insights gained here may guide extensions to more complex configurations.

Let us briefly review previous results within this model that are relevant to the present study. The existence of self-trapped bright soliton solutions above a certain threshold atom number was demonstrated in \cite{petrov2016ultradilute}. As the number of atoms increases, the soliton profiles evolve from bell-shaped to flat-top \cite{astrakharchik2018dynamics,pathak2022dynamics,huang2022binary}. Solutions in the flat-top regime can be interpreted as a liquid-like phase \cite{michinel2002} of the atomic cloud, where an almost constant-density fluid forms droplets with well-defined surface tension and obeys a variant of the Young-Laplace equation \cite{paredes2024variational}. The model also supports self-trapped vortex solitons, as first demonstrated in \cite{li2018two} and further explored in a range of works including \cite{zhang2019semidiscrete,caldara2022vortices,liu2022vortex,zhang2023collisional,salgueiro2024stability}.

The symmetric 2D model has also served to investigate various dynamical aspects of QDs, including collisional processes \cite{hu2022collisional}, turbulent regimes \cite{jha2025energy}, and the stability characteristics of nonlinear excitations \cite{bougas2024stability}.

This work builds naturally upon our previous study in Ref.~\cite{paredes2025traveling}, where we identified dark traveling wave eigenstates propagating within an infinite QD background. However, physically relevant QDs are necessarily finite in extent. The present study aims to analyze the behavior of such excitations when embedded in large but finite bright solitons. A defining feature of QDs is the interplay between attractive interspecies interactions at low densities and repulsive intraspecies interactions at higher densities, which enables the self-sustained coexistence of both dark and bright solitary waves within the same physical system. This raises the central question of this work: how do dark excitations behave when they reach the boundary of the hosting bright soliton? In what follows, we explore the rich phenomenology that can emerge in this regime, focusing on the behavior of traveling dark excitations as they approach the boundary of a bright droplet, their dynamic generation near the edge via collisions between distinct QDs, and other related processes.

\section{Mathematical formalism and setup}
\label{sec:math}

Throughout this paper, we work with the generalization of the GPE that describes two-dimensional symmetric BEC mixtures in the presence of the LHY correction, as derived in \cite{petrov2016ultradilute}. For convenience, we use the following adimensional form of the equation, which can be obtained from the dimensional model via rescaling \cite{paredes2025traveling}:
\begin{equation}
i\partial_{ t}  \Psi = -  \nabla^2  \Psi +  |\Psi|^2\ln  (|\Psi|^2)  \Psi .
\label{eqadim}
\end{equation}
This equation supports stable bright soliton solutions \cite{petrov2016ultradilute} of the form:
\begin{equation}
\Psi_{sol} = e^{-i\mu t} \psi_{sol}(r),
\label{psisol}
\end{equation}
where $\psi_{sol}(r)$ is a real, monotonically decreasing function, and $(r,\theta)$ denote polar coordinates in the plane. The number of atoms is proportional to the norm of the wavefunction:
\begin{equation}
N=\int |\Psi_{sol}|^2 d^2{\bf r} = 2\pi \int \psi_{sol}(r)^2 r\, dr.
\label{Norm}
\end{equation}
Soliton solutions exist for $0 > \mu > \mu_{cr}$, where the critical chemical potential is given by:
\begin{equation}
\mu_{cr} = -\frac{1}{2\sqrt{e}},
\end{equation}
see, e.g., \cite{paredes2024variational}. As $\mu \to \mu_{cr}$, the number of atoms diverges, $\lim_{\mu\to\mu_{cr}} N = \infty$, and the profile approaches a flat-top shape: $\psi_{sol}(r) \approx \psi_{cr}$ for $r < R$ and $\psi_{sol}(r) \approx 0$ for $r > R$, where $R$ is the droplet radius and $N \approx \pi R^2 \psi_{cr}^2$. The critical amplitude is:
\begin{equation}
\psi_{cr} = e^{-\frac14}.
\end{equation}
A good approximation for the soliton profile in this flat-top limit is given by \cite{paredes2024variational}:
\begin{equation}
\psi_{sol}(r) \approx \psi_{cr} \sqrt{\frac{1+\tanh (\kappa (R-r))}{2}}, \quad   (\mu \to \mu_{cr},\, R \gg 1),
\label{psi_var}
\end{equation}
where $\kappa = \frac{\sqrt{12-\pi^2}}{\sqrt{6} \,e^{1/4}} \approx 0.4641$. This model also admits stable self-trapped solitons with angular momentum \cite{li2018two} of the form:
\begin{equation}
\Psi_{vort} = e^{-i\mu t} e^{i \ell \theta} \psi_{vort}(r),
\label{psivort}
\end{equation}
where $\ell \in \mathbb{Z}$ is the vorticity. These solutions resemble bright rings with a phase singularity of order $\ell$ at their center. 
They also exhibit a flat-top limit as $\mu \to \mu_{cr}$ and $N \to \infty$, where the profile can be approximated by a
 generalization of (\ref{psi_var}); see \cite{paredes2024variational} for details.

Another class of eigenstate solutions of Eq.~(\ref{eqadim}) was presented in \cite{paredes2025traveling}, following similar constructions for other 
nonlinear wave equations \cite{jones1982motions,paredes2014coherent,chiron2016travelling}. These solutions describe traveling waves in which a localized density dip (a ``bubble'') 
moves at constant velocity:
\begin{equation}
\Psi_{bubble} = e^{-i \mu_{cr}t} \psi_{bubble}(x-Ut,y),
\label{psibubble}
\end{equation}
where $(x,y)$ are Cartesian coordinates. Equation~(\ref{psibubble}) represents form-preserving excitations over the critical background: $\psi_{bubble}$ is a complex function satisfying the boundary condition $\lim_{r \to \infty} \psi_{bubble} = \psi_{cr}$. The ansatz of Eq. (\ref{psibubble})
 is written for motion along the $x$-axis, but can be rotated to any direction in the plane.

We use the term ``bubble'' \cite{berloff2004vortex,zeng2021flat,katsimiga2023} to describe localized regions where the atomic density is reduced relative to the background value $\psi_{cr}^2$. These solutions exist for velocities $0 < U < U_0$, where $U_0 = e^{-1/4}$ is the long-wavelength speed of sound in the fluid. At low velocities, the bubbles consist of vortex-antivortex pairs, while at higher velocities they become rarefaction pulses without phase singularities \cite{paredes2025traveling}.

Although the bubbles described by Eq.~(\ref{psibubble}) are exact solutions only in an infinite (unphysical) critical background, it is natural to ask whether they can behave as excitations within large, near-critical, flat-top solitons or vortices. We address this question in the following sections, primarily through numerical simulations of Eq.~(\ref{eqadim}). For our computations, we use a standard split-step pseudo-spectral method; see \cite{figueiras2018open} for an introductory overview.

We will show that the density dips described by Eq.~(\ref{psibubble}) propagate largely undistorted inside the QDs defined by Eqs.~(\ref{psisol}) or (\ref{psivort}) until they approach the boundary of the droplet, where non-trivial bubble-boundary interactions occur.

Our initial conditions are constructed from Eqs.~(\ref{psisol}) and (\ref{psibubble}). Specifically:
\begin{equation}
\Psi_0(t=0) = \frac{1}{\psi_{cr}} \, \psi_{sol}(r) \psi_{bubble}(x_0,y_0),
\label{Psi0}
\end{equation}
where the profiles $\psi_{sol}$ and $\psi_{bubble}$ are computed numerically. This ansatz corresponds to a soliton centered at the origin, and, within the soliton, a bubble centered at $(x_0, y_0)$. This configuration carries nonzero momentum along the $x$-direction:
\begin{equation}
p_x = \frac{1}{2i} \int \left[ \Psi^* \partial_x \Psi -  \Psi \partial_x \Psi^* \right] dx\,dy.
\label{px}
\end{equation}
Due to the Galilean symmetry of the GPE (\ref{eqadim}), momentum is conserved, and the center of mass of the configuration moves at velocity $v_x = 2p_x/N$. Moreover, Galilean invariance implies that if $\Psi$ is a solution of Eq.~(\ref{eqadim}), then so is $e^{\frac{i}{2}v_x x} \Psi$, representing a solution boosted at velocity $v_x$.

Therefore, from Eq.~(\ref{Psi0}) we can construct an ansatz for which the center of mass remains at rest:
\begin{equation}
\Psi(t=0) = e^{-i\,\frac{p_{x,0}}{N}x} \Psi_0(t=0),
\label{Psi0bis}
\end{equation}
where $p_{x,0}$ is computed by inserting Eq.~(\ref{Psi0}) into Eq.~(\ref{px}). While the dynamics obtained with initial conditions given in Eqs.~(\ref{Psi0}) and (\ref{Psi0bis}) 
are physically equivalent, using Eq.~(\ref{Psi0bis}) helps avoid the atomic cloud drifting toward the computational boundary, which can cause numerical issues, particularly in long-time simulations.

\section{Travelling waves interacting with the boundary of a soliton}
\label{sec:bdy}

As outlined above, we start our numerical analysis by simulating the evolution of different bubble solutions within a flat-top soliton.
In the rest of this section, we take as the hosting soliton the solution of the form (\ref{psisol}) that can be approximated by (\ref{psi_var}) with $R=132$.
We then construct the initial conditions from Eq. (\ref{Psi0bis}), as explained in detail in section \ref{sec:math}.
The qualitative behaviour for flat-top solitons of other sizes is equivalent. 

\begin{figure}[h!]
\begin{center}
\includegraphics[width=\columnwidth]{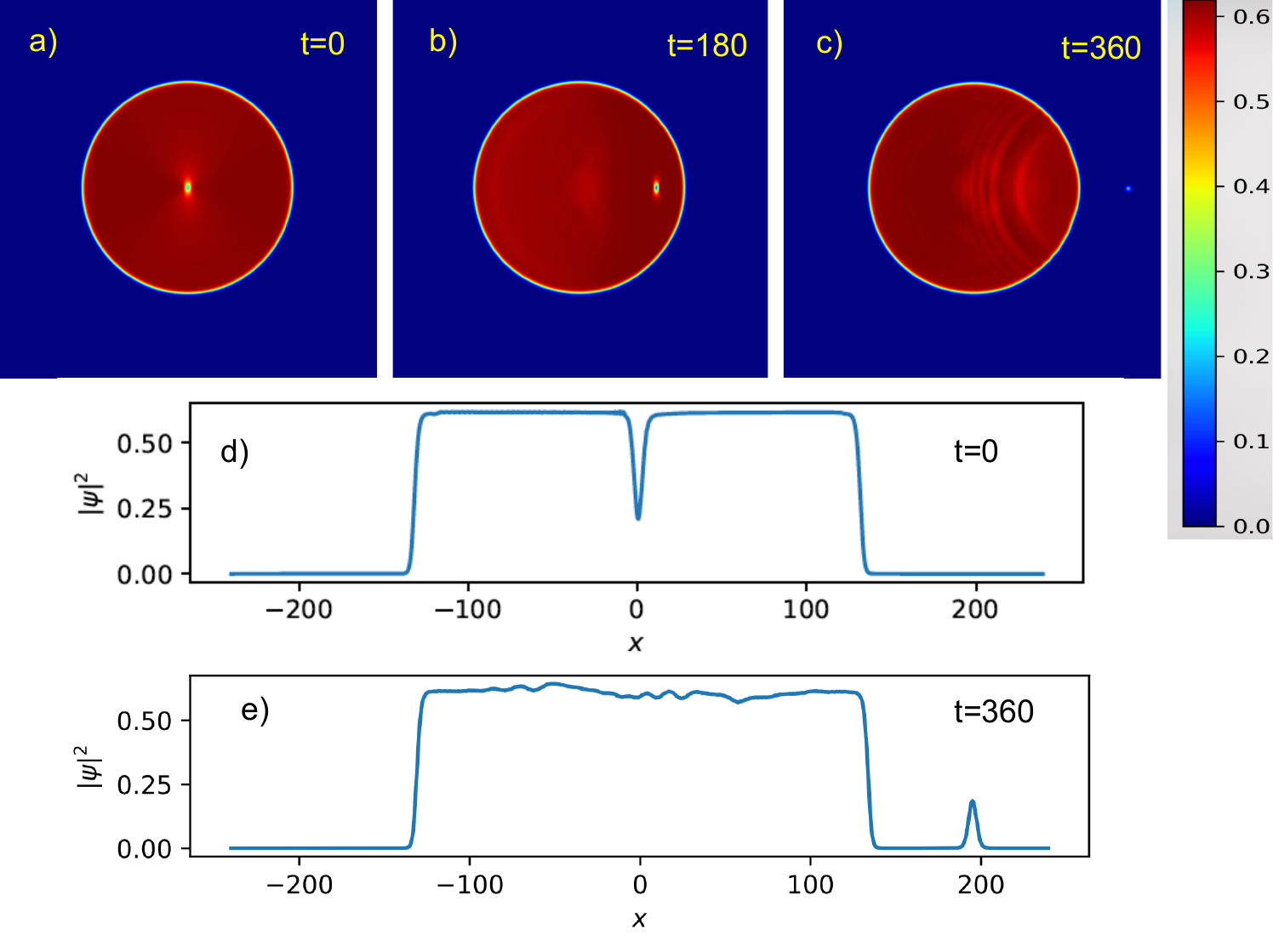}
\end{center}
\caption{A faint rarefaction pulse produces an outgoing bell-shaped soliton when interacting with the boundary of the hosting soliton. 
Panels a)-c) show a colormap of $|\Psi(x,y)|^2$ at different times. The size of each window in the $x-y$ plane is $480\times 480$. Panels d) and e) are cuts
along the $x$ direction, namely $|\Psi(x,0)|^2$.}
\label{fig1}
\end{figure}

Let us first compute the evolution of a rarefaction pulse whose minimum density is well above zero. The result is displayed in Fig. \ref{fig1}, that
shows that, when the bubble reaches the boundary, a bell-shaped soliton is emitted, leaving the flat top soliton in
an excited stated, with ripples propagating inside it. Videos for this and the rest of simulations described in this paper are provided
 in the supplemental material \cite{supplemental}.

\begin{figure}[h!]
\begin{center}
\includegraphics[width=\columnwidth]{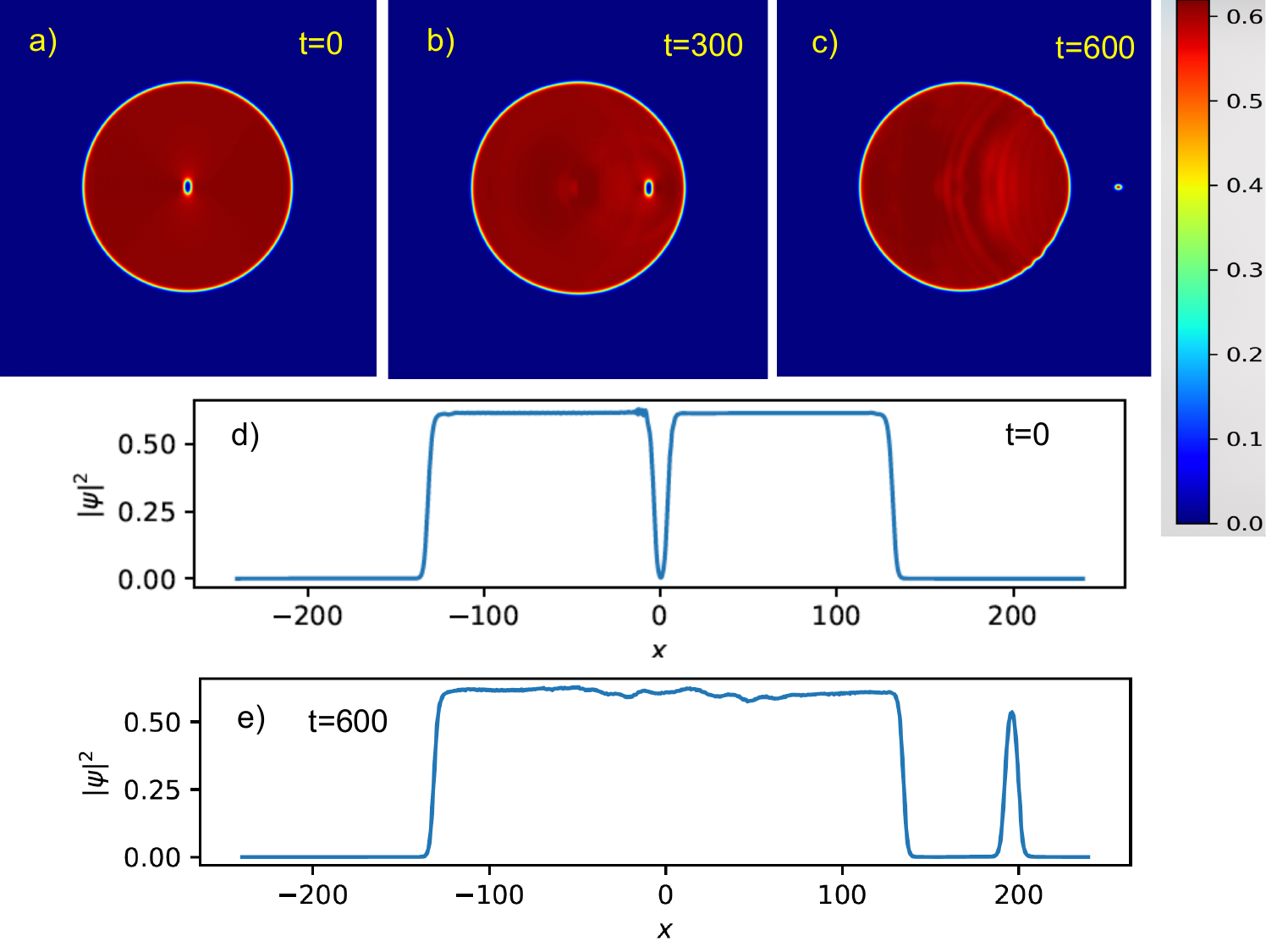}
\end{center}
\caption{A rarefaction pulse close to the transition to vortex-antivortex pair  produces an outgoing bell-shaped soliton
close to the transition to flat-top profiles.
Panels a)-c) show a colormap of $|\Psi(x,y)|^2$ at different times, with the same convention of Fig. \ref{fig1}. Panels d) and e) are $|\Psi(x,0)|^2$ cuts.}
\label{fig2}
\end{figure}

As we increase the energy of the rarefaction pulse or, equivalently, decrease its velocity (see \cite{paredes2025traveling} for a detailed discussion on the rarefaction pulses), the
emitted soliton becomes larger. Figure \ref{fig2} shows a case in which the minimum density of the rarefaction pulse is close to zero, and therefore it is close to the transition
between rarefaction pulses and vortex-antivortex pairs. The outgoing soliton is still bell-shaped, but its maximum density is near $\psi_{cr}^2$, showing that it is
close to the transition between bell-shaped and flat-top solitons.

\begin{figure}[h!]
\begin{center}
\includegraphics[width=\columnwidth]{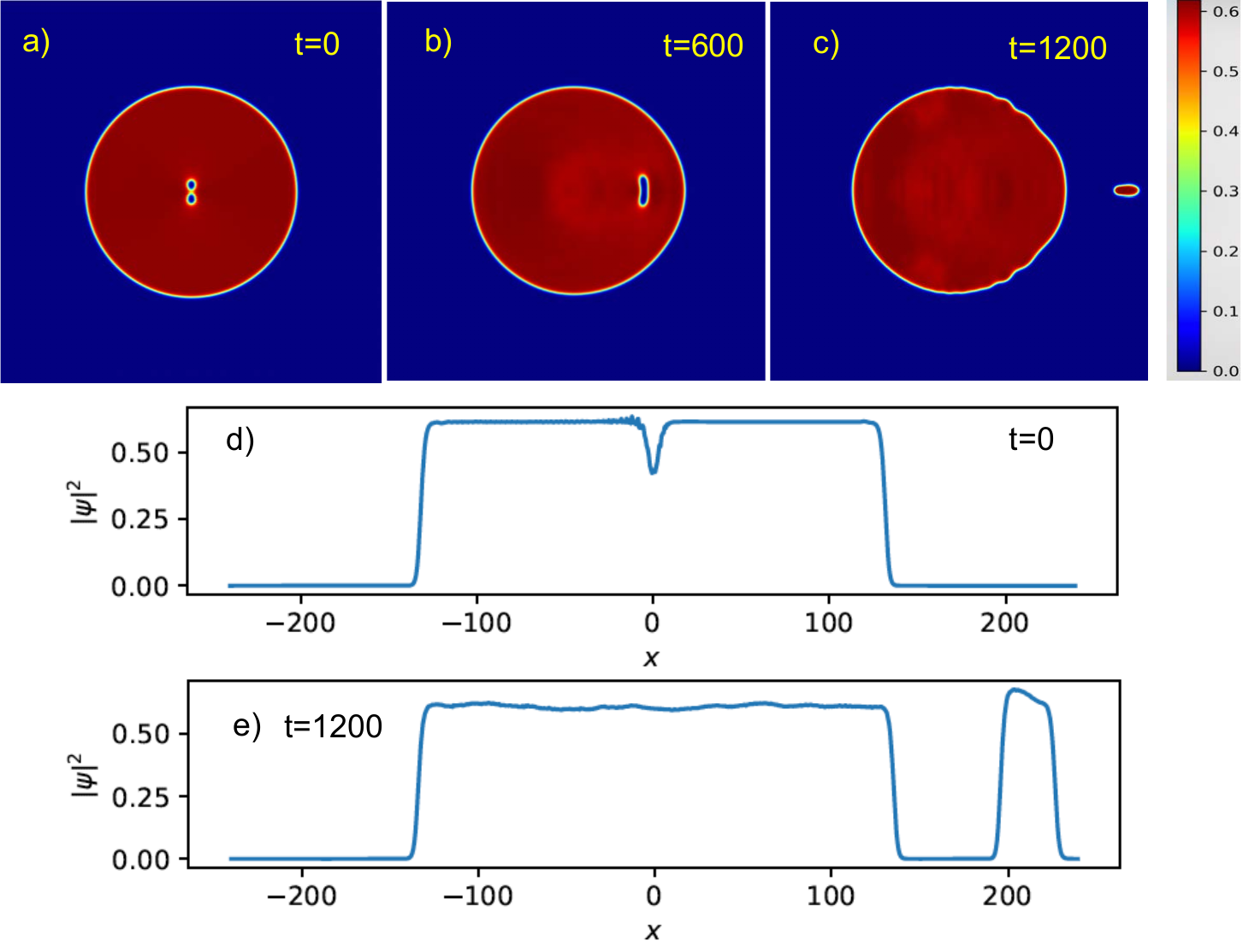}
\end{center}
\caption{A vortex-antivortex pair with velocity $U \approx U_{sc}$ produces an outgoing flat-top soliton, while the initial hosting soliton gets strongly
perturbed in the process.
Panels a)-c) show a colormap of $|\Psi(x,y)|^2$ at different times, with the same convention of Fig. \ref{fig1}. Panels d) and e) are $|\Psi(x,0)|^2$ cuts.}
\label{fig3}
\end{figure}

As shown in \cite{paredes2025traveling}, there are rarefaction pulses without phase singularities for a range of velocities $U_{rp}<U<U_0$. 
When we look for solutions of the form of Eq.~(\ref{psibubble}) with $U<U_{rp}$, two phase singularities appear --- namely, a vortex-antivortex pair. 
For a certain range of velocities $U_{sc}<U<U_{rp}$, both singularities lie within the same density minimum, forming an elongated trough-like region
 in the atomic density distribution. When this kind of excitation reaches the boundary, it results in the emission of a flat-top soliton whose size is 
 comparable to that of the trough. Heuristically, one can think of the elongated dip as splitting the flat-top soliton into two fragments --- a larger one 
 and a smaller one --- which then separate from each other. This behavior is illustrated in Fig.~\ref{fig3}, where the simulation is performed with a 
 vortex-antivortex pair moving at a velocity $U \approx U_{sc}$.

\begin{figure}[h!]
\begin{center}
\includegraphics[width=\columnwidth]{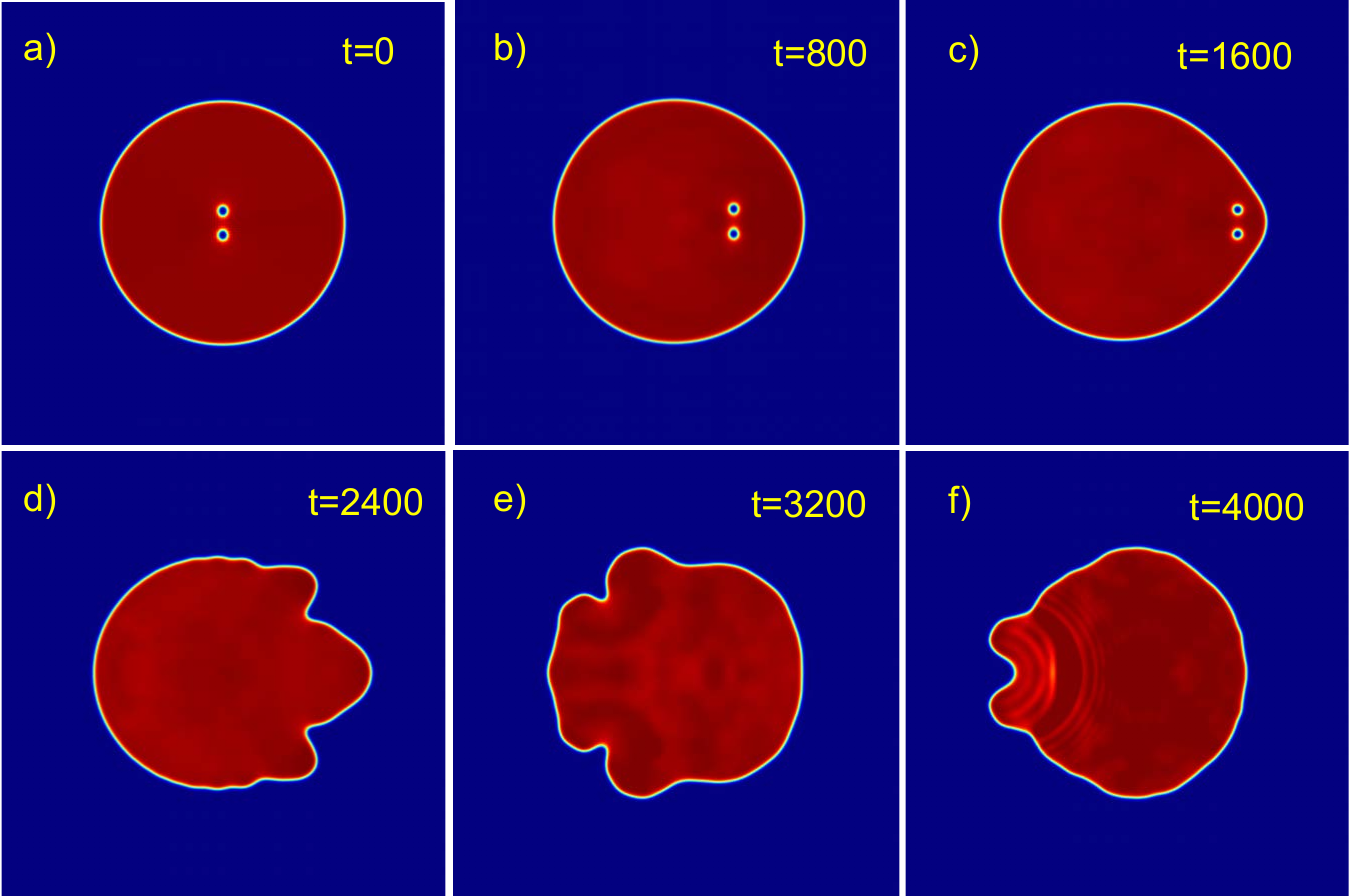}
\end{center}
\caption{A vortex-antivortex pair with velocity $U < U_{sc}$ disappears when reaching the edge of the flat-top soliton, which gets strongly
perturbed in the process.
All panels show a colormap of $|\Psi(x,y)|^2$ at different times, with the same convention of Fig. \ref{fig1}.}
\label{fig4}
\end{figure}

Let us now consider vortices with $U<U_{sc}$ such that the vortex and antivortex lie in separate vortex cores. In this case, our numerical simulations show 
that there is no soliton emission. The density dips around the vortex and antivortex eventually reach the edge of the hosting soliton, and the phase singularities 
are expelled from it, leaving the soliton in a rather excited state while some atoms escape without forming new self-trapped states. Heuristically, one can 
understand the absence of soliton emission by noting that the vortex and antivortex, being separated, do not generate two disconnected regions; instead, they 
remain linked by a  bridge of non-vanishing density, as shown in Fig.~\ref{fig4}.

\begin{figure}[h!]
\begin{center}
\includegraphics[width=\columnwidth]{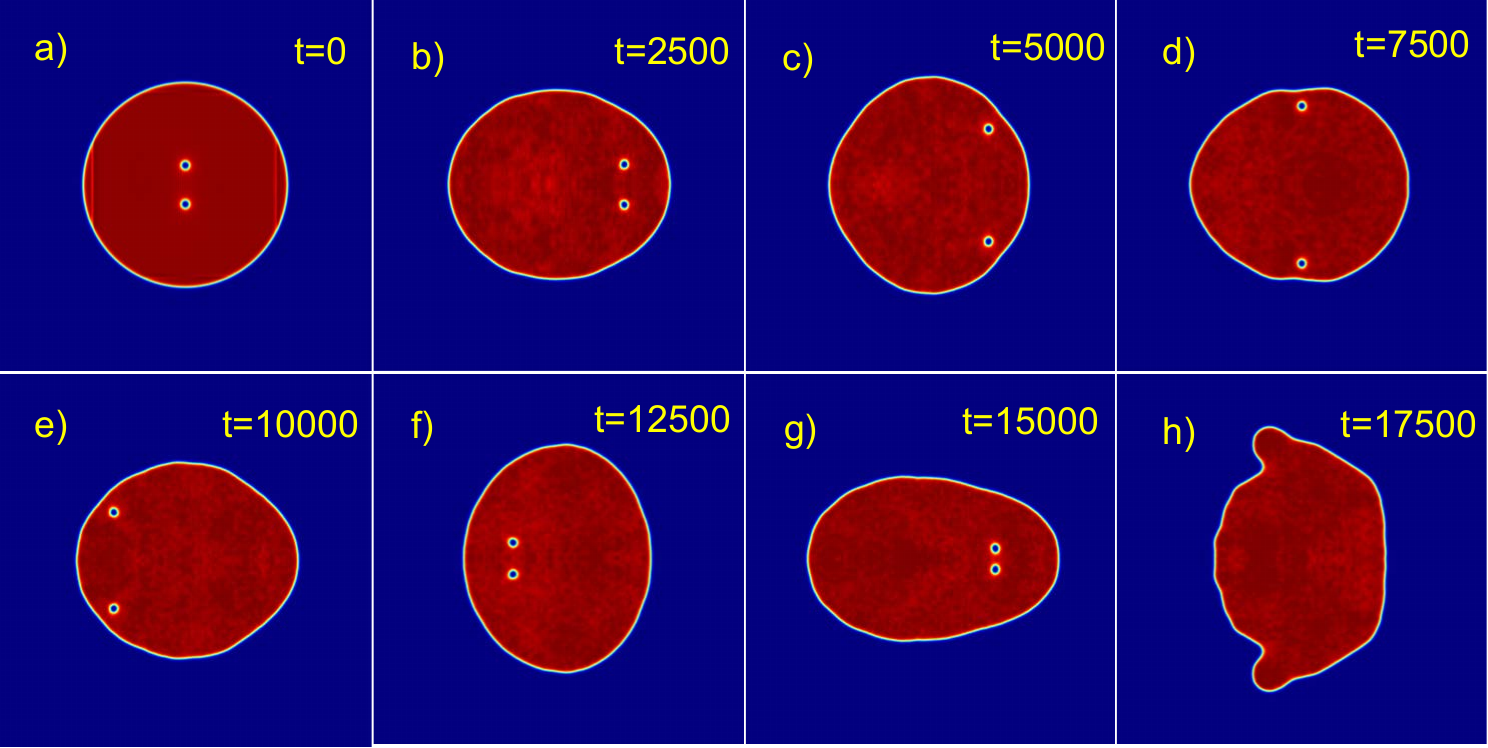}
\end{center}
\caption{A widely separated vortex-antivortex pair becomes split upon reaching the edge of the flat-top soliton. Both the vortex and the antivortex start moving 
backward due to their interaction with the boundary. Eventually, they meet again and couple, resuming their joint motion to the right. Finally, they reach the edge 
of the soliton once more, where the phase singularities are expelled.
All panels show a colormap of $|\Psi(x,y)|^2$ at different times, with the same convention of Fig. \ref{fig1}.}
\label{fig5}
\end{figure}

If we further increase the separation between the vortex and antivortex, thus reducing their velocity, a new qualitative behavior emerges. The vortex and
 antivortex move apart and begin propagating backward near the edge of the droplet. Eventually, due to the droplet being far from equilibrium, 
 the outer boundary can reach the vortex core, effectively removing the singularity from the interior of the flat-top region. This process is illustrated in 
 Fig.~\ref{fig5}. In that particular case, the vortex and antivortex travel along the entire semicircular boundary and eventually recombine into a rightward-moving
  vortex-antivortex pair, which splits once more and is ultimately destroyed. Interestingly, if the simulation in Fig.~\ref{fig5} is continued, 
  the dynamical evolution of the excited droplet ultimately gives rise to a rarefaction pulse that emits a small soliton (not shown in the figure but available 
  in the supplemental video \cite{supplemental}). Thus, the simulation in Fig.~\ref{fig5} encapsulates various qualitative behaviors that characterize the dynamics of traveling waves
  in quantum droplets.

We can provide a very rough approximation for the vortex-antivortex distance that marks the transition between the direct expulsion of the phase singularities
 (Fig.~\ref{fig4}) and the separation of the vortex-antivortex pair along the droplet boundary (Fig.~\ref{fig5}). To do so, we must first explain why vortices tend 
 to split and rotate along the edge of the droplet. It is well known that, due to phase boundary conditions, a vortex located at a distance $d$ from the center of a 
 circular domain of radius $R$ behaves as if it were being influenced by an image antivortex located at a distance $\frac{R^2}{d}$ from the center; 
 see, for example, \cite{havelock1931lii,groszek2018motion,paredes2023polygons}. A vortex of charge 1 at radial position $d$ thus experiences the effect of
  an antivortex located at a distance $\frac{R^2}{d} - d \approx 2(R-d)$, assuming $R-d \ll d$. As a result, it is dragged along the fluid 
  with a velocity approximately given by $\frac{1}{R-d}$. 

This velocity should be compared with that induced by the real antivortex of the pair, which for a vortex-antivortex separation $L$ is approximately 
$\frac{2}{L}$. When $\frac{1}{R-d} > \frac{2}{L}$, the boundary effect becomes dominant, and the pair begins to split. For the vortex and antivortex to survive 
this process, the splitting should occur not too close to the boundary --- say, at a distance $R-d \approx 4 R_{\textrm{vortex core}}$, where the vortex core 
radius can be estimated by numerically solving $\psi_{\textrm{vort}}(r)$ from Eq.~(\ref{psivort}) with $\ell = 1$. In our model, this 
gives $R_{\textrm{vortex core}} \approx 5$ \cite{paredes2024variational}. Substituting this into the inequality above, we estimate that the vortex and antivortex 
will tend to split and propagate near the droplet boundary when $L > 40$. 
For comparison, in Fig.~\ref{fig4} we have $L \approx 25$, while in Fig.~\ref{fig5} we have $L \approx 50$. Clearly, this argument is not rigorous or quantitative, but it
 provides a heuristic understanding of the dynamics underlying the distinct behaviors observed in the numerical simulations.

\begin{figure}[h!]
\begin{center}
\includegraphics[width=\columnwidth]{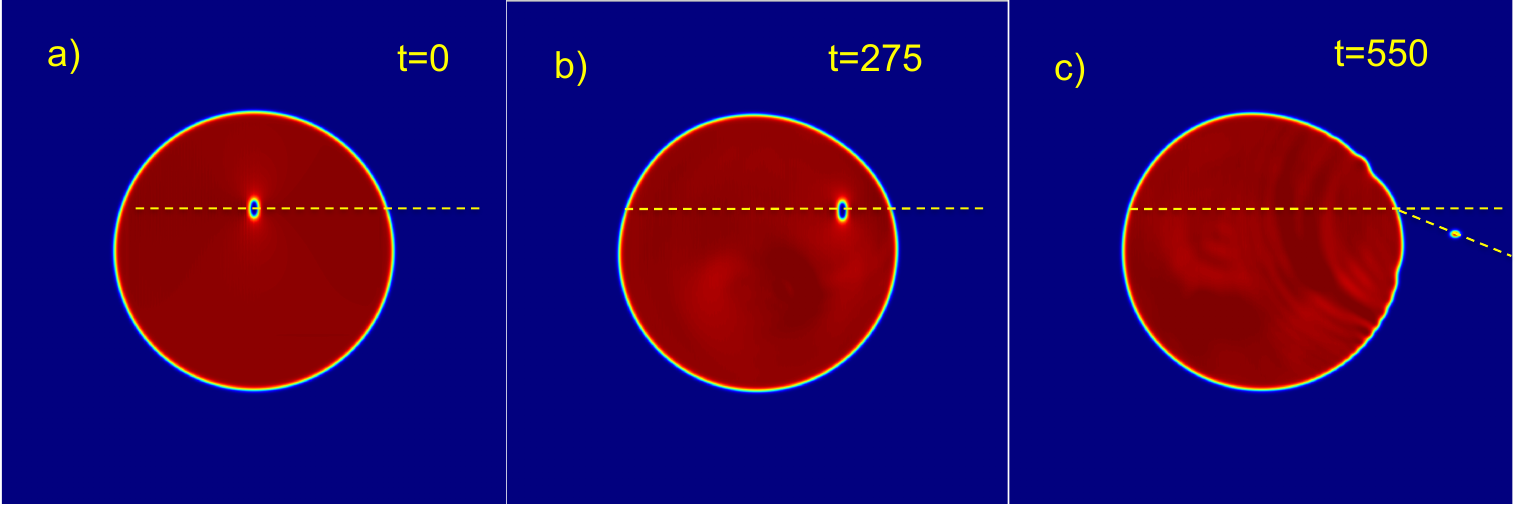}
\end{center}
\caption{The rarefaction pulse of Fig. \ref{fig2} impinges non-perpendicularly on the boundary of the droplet. It generates an outgoing soliton that moves
oblique to the original trajectory. 
Dashed yellow lines have been introduced to highlight this fact. All panels show a colormap of $|\Psi(x,y)|^2$ at different times, with the same convention of Fig. \ref{fig1}.}
\label{fig6}
\end{figure}

Up to this point, we have only considered diametral trajectories. It is natural to also wonder what happens in the case in which the  bubble does not reach the 
boundary perpendicularly. Our numerical simulations show that the qualitative behaviors remain the same, with the particularity that the outgoing soliton
moves in a different direction compared to the initial travelling wave, in a sort of nonlinear refraction phenomenon. We illustrate that behavior with 
an example in Fig. \ref{fig6}.

\section{Travelling waves interacting with the boundary of a vortex soliton}
\label{sec:vortex}

A natural question is how the dynamics described in the previous section are modified when the hosting quantum droplet carries vorticity. The goal of this 
section is to briefly address this issue. It turns out that the interaction of rarefaction pulses and vortex-antivortex pairs with the droplet boundary remains 
essentially the same as that discussed in Section~\ref{sec:bdy}. The main difference is that any travelling wave is now dragged by the background 
flow associated with the droplet's angular momentum, thereby modifying its trajectory.

\begin{figure}[h!]
\begin{center}
\includegraphics[width=\columnwidth]{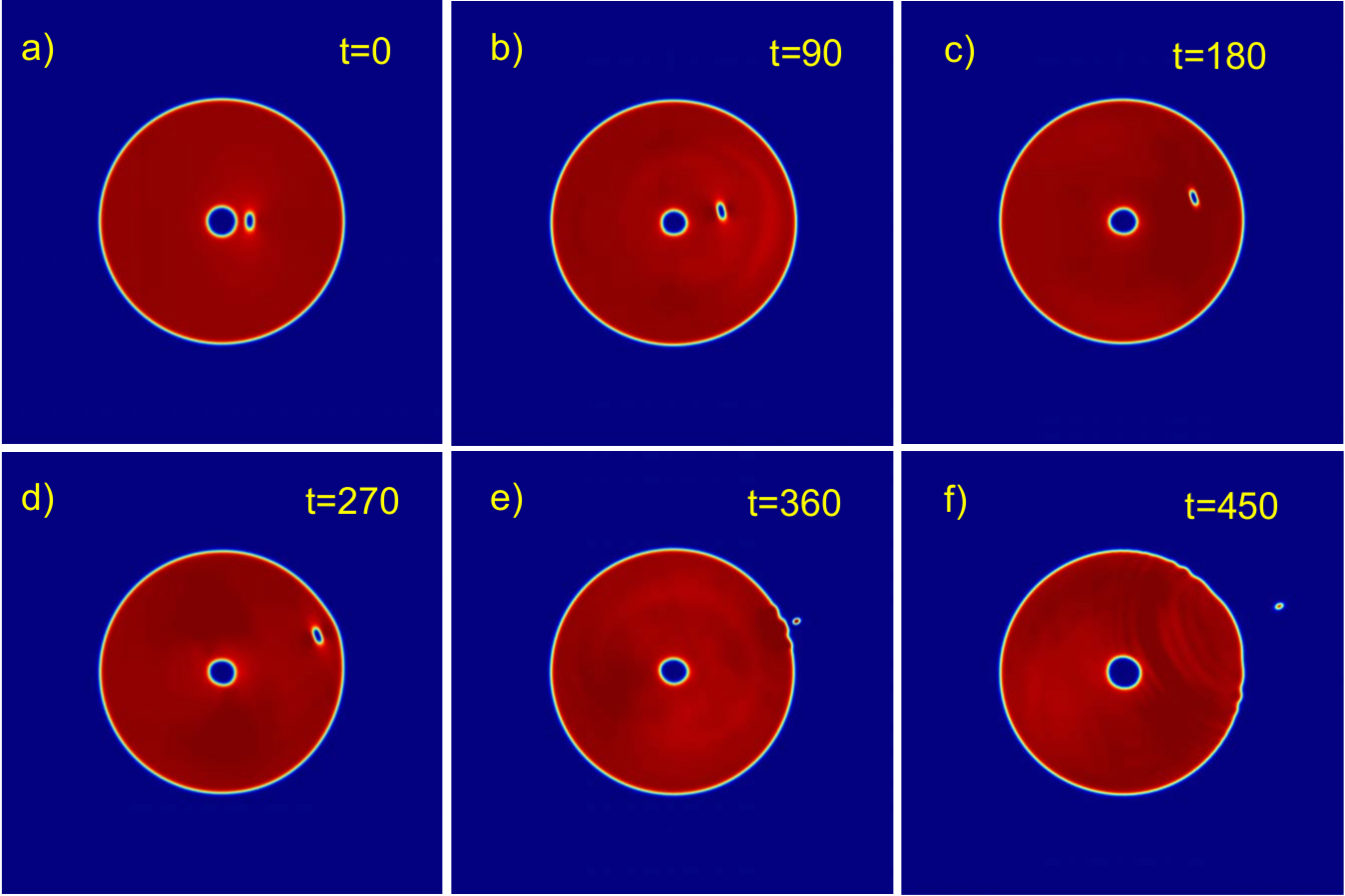}
\end{center}
\caption{The rarefaction pulse from Fig.~\ref{fig2} is placed inside a large vortex soliton with vorticity $\ell = 2$ and outer radius $R \approx 132$. 
The excitation moves under its own velocity while also being dragged by the background rotation of the fluid. Eventually, it reaches the boundary 
and triggers the emission of a bright soliton that propagates freely.
All panels show a colormap of $|\Psi(x,y)|^2$ at different times, with the same convention of Fig. \ref{fig1}.}
\label{fig7}
\end{figure}

A first illustration of this effect is shown in Fig.~\ref{fig7}, where we place a rarefaction pulse near the inner boundary of an $\ell = 2$ bright vortex soliton. 
The pulse is initially oriented to move radially outward. However, the counterclockwise rotation of the background fluid causes its trajectory to curve, as seen in the figure. 
Had we instead used a clockwise-rotating $\ell = -2$ flat-top vortex, the pulse would have curved in the opposite direction. In either case, once the pulse reaches the boundary, it triggers the emission of a bright soliton, which then propagates freely --- similarly to what is shown in Fig.~\ref{fig2}.

\begin{figure}[h!]
\begin{center}
\includegraphics[width=\columnwidth]{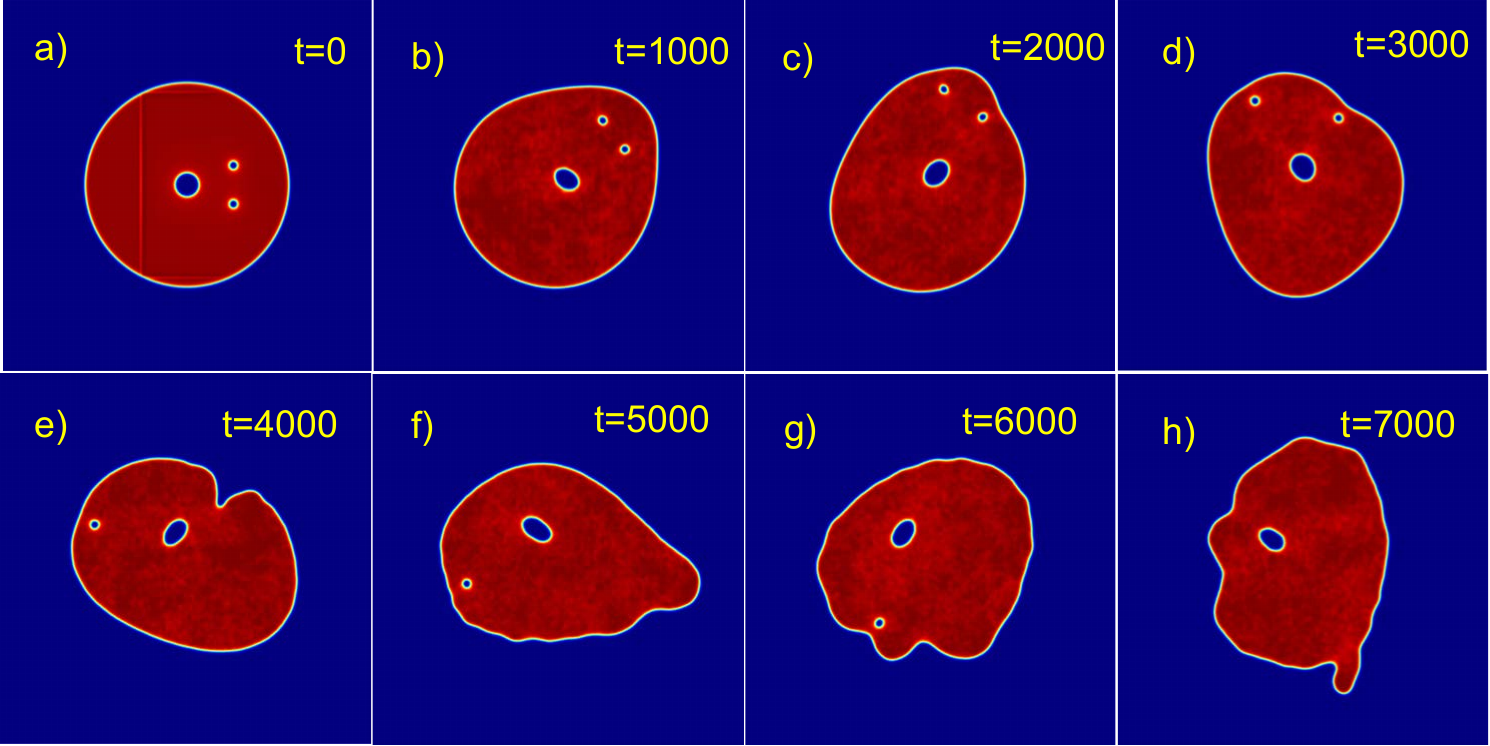}
\end{center}
\caption{The vortex-antivortex pair from Fig.~\ref{fig5} is now placed inside the same vortex soliton used in Fig.~\ref{fig7}, with vorticity $\ell = 2$. 
The pair moves under its own velocity while also being advected by the rotation of the background fluid. As it approaches the boundary, the two singularities 
split: the vortex follows a curved trajectory along the edge, while the antivortex moves more slowly. When a deformation of the hosting soliton reaches one of 
the cores, the corresponding phase singularity is expelled. All panels show a colormap of $|\Psi(x,y)|^2$ at different times, using the same conventions as in Fig.~\ref{fig1}.}
\label{fig8}
\end{figure}

Figure~\ref{fig8} shows another representative case, where a vortex-antivortex pair is initially embedded within the $\ell = 2$ bright vortex soliton. 
As in the previous example, the excitation moves under its own velocity while also being advected by the counterclockwise rotation of the background fluid.

Compared to the rarefaction pulse of Fig.~\ref{fig7}, the pair rotates more significantly before reaching the boundary. This is because its intrinsic propagation speed is 
lower, while the rotational dragging remains the same. As the pair approaches the edge, it splits: the vortex is pushed counterclockwise along the boundary by the
 combined effects of its associated image antivortex (induced by the droplet's finite size) and the central $\ell = 2$ vortex. The antivortex, on the other hand, moves 
 much more slowly, since the two contributing effects act in opposite directions.
 
Eventually, a deformation of the boundary reaches the density dip surrounding each of the singularities, leading to its expulsion from the droplet.

Another qualitative effect that can arise in this type of simulation is a strong interaction between the travelling waves and the central vortex core, 
which can substantially deform or destabilize the soliton. For example, if in Fig.~\ref{fig8} the vortex-antivortex pair had been initialized closer to the center, 
one of the phase singularities could merge with the $\ell = 2$ core. In that case, the density dips surrounding the singularities can connect, allowing the redistribution
 of vorticity within the droplet. While the total winding number remains unchanged as long as all singularities remain inside the soliton, the resulting configuration
  no longer resembles a well-defined central vortex with separate excitations. This illustrates that the inserted
   excitations can strongly reshape the internal structure of the droplet, beyond the regime of small perturbations.

\section{Travelling waves from droplet-droplet collisions}
\label{sec:collisions}

  In the previous sections, we considered initial conditions in which dark travelling waves were embedded within a host soliton or vortex, 
  showing that one possible outcome is the emission of a soliton.
Here, we demonstrate that the reverse process is also possible: the collision of a soliton or vortex with a smaller soliton can lead to the excitation of 
a rarefaction pulse or the formation of a vortex-antivortex pair. This process represents a form of coherent cavitation, as originally discussed
 in \cite{paredes2014coherent} in a different context.
We illustrate this through three representative examples, all involving a large vortex soliton and a smaller bright soliton moving
 toward each other as the initial condition. Although similar effects can result from collisions between solitons of different sizes, 
 we focus on cavitation for a flat-top vortex, which, to the best of our knowledge, has not yet been reported in any physical model. In all simulations, 
 we use the same $\ell = 2$ vortex soliton introduced in Section~\ref{sec:vortex}; however, the qualitative behaviour remains the same for 
 other values of $\ell$ and different flat-top plateau sizes.

\begin{figure}[h!]
\begin{center}
\includegraphics[width=\columnwidth]{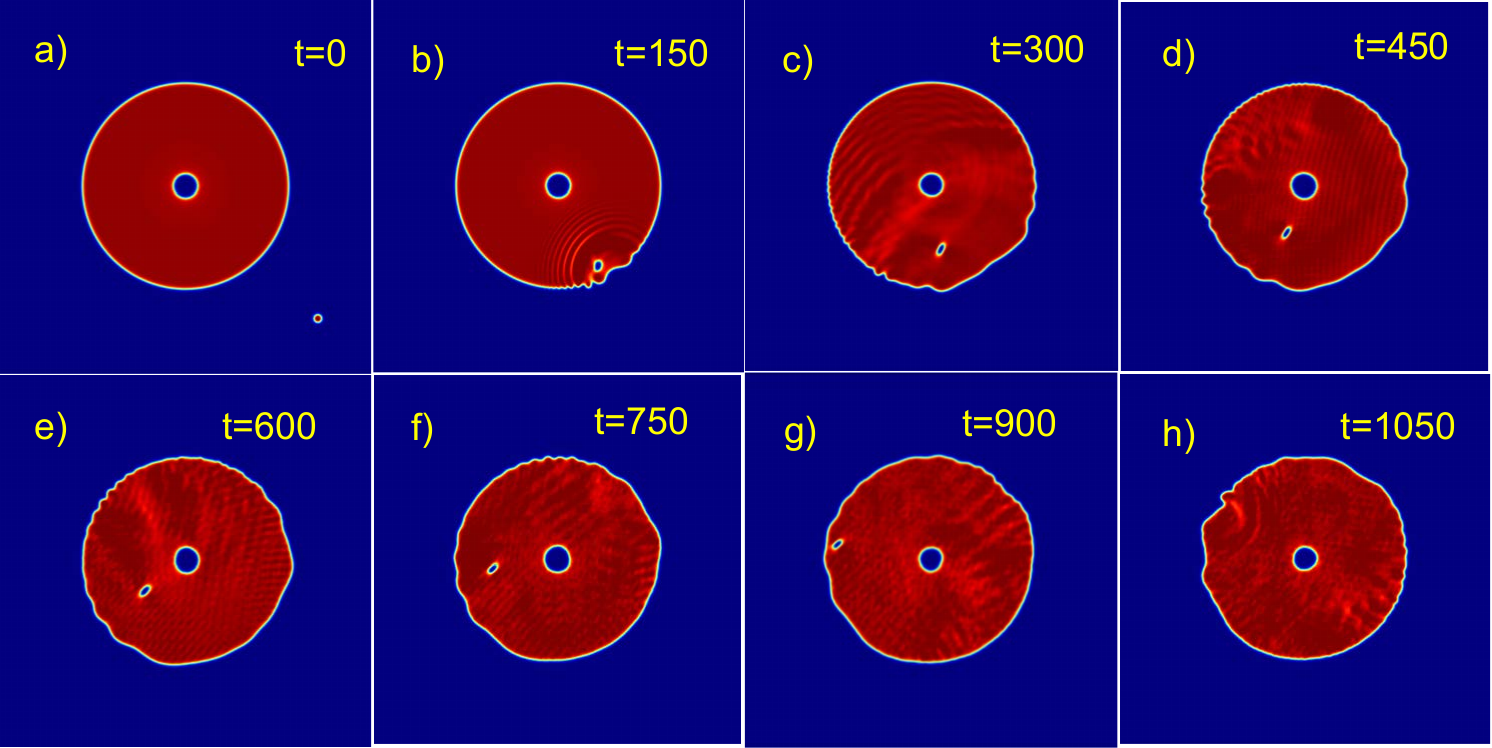}
\end{center}
\caption{Soliton-vortex collision generating a rarefaction pulse moving counter to the  rotating fluid of the vortex.
 All panels show a colormap of $|\Psi(x,y)|^2$ at different times, using the same conventions as in Fig.~\ref{fig1}.}
\label{fig9}
\end{figure}

Figure~\ref{fig9} shows how a soliton generates a rarefaction pulse upon colliding with the flat-top vortex soliton, which subsequently propagates inside the vortex.
We observe that this outcome depends on the relative phase between the initially separated eigenstates. With different relative phases, 
the soliton may be mostly absorbed or even bounce back from the vortex boundary.
The result also depends on the relative velocity: rarefaction pulses are only formed above a certain velocity threshold.

\begin{figure}[h!]
\begin{center}
\includegraphics[width=\columnwidth]{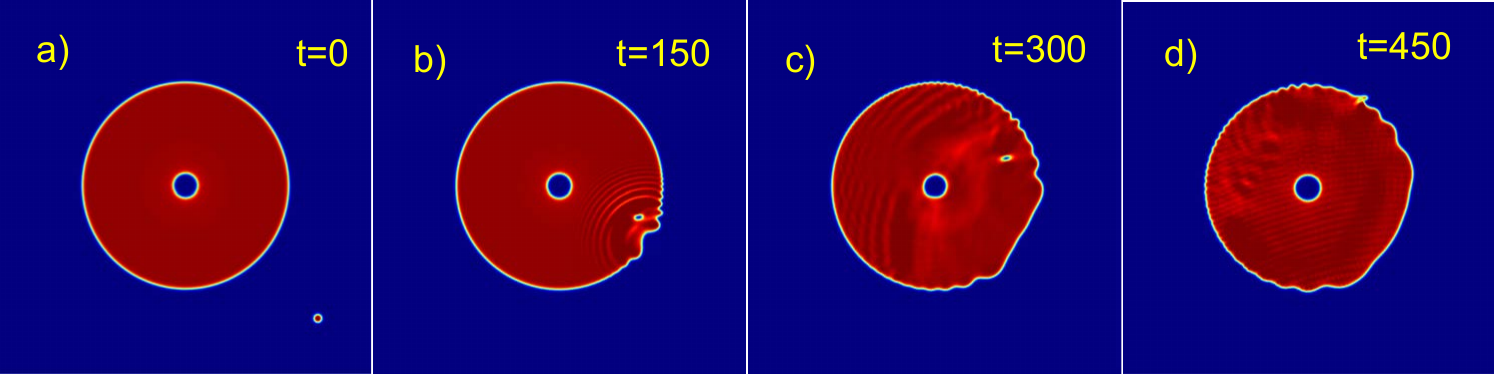}
\end{center}
\caption{A soliton-vortex collision generates a rarefaction pulse that moves along with the rotating fluid of the vortex.
 All panels show a colormap of $|\Psi(x,y)|^2$ at different times, using the same conventions as in Fig.~\ref{fig1}.}
\label{fig10}
\end{figure}

Figure~\ref{fig10} presents a second example using the same initial eigenstates as in Fig.~\ref{fig9}, but with a different collision orientation. 
In this case, the rarefaction pulse propagates along the counterclockwise rotation of the vortex fluid. As a result, it moves faster than the pulse 
in Fig.~\ref{fig9}, as can be appreciated by comparing the figures. This provides a clear illustration of how the properties of
 the background affect the trajectories of dark pulses.

\begin{figure}[h!]
\begin{center}
\includegraphics[width=\columnwidth]{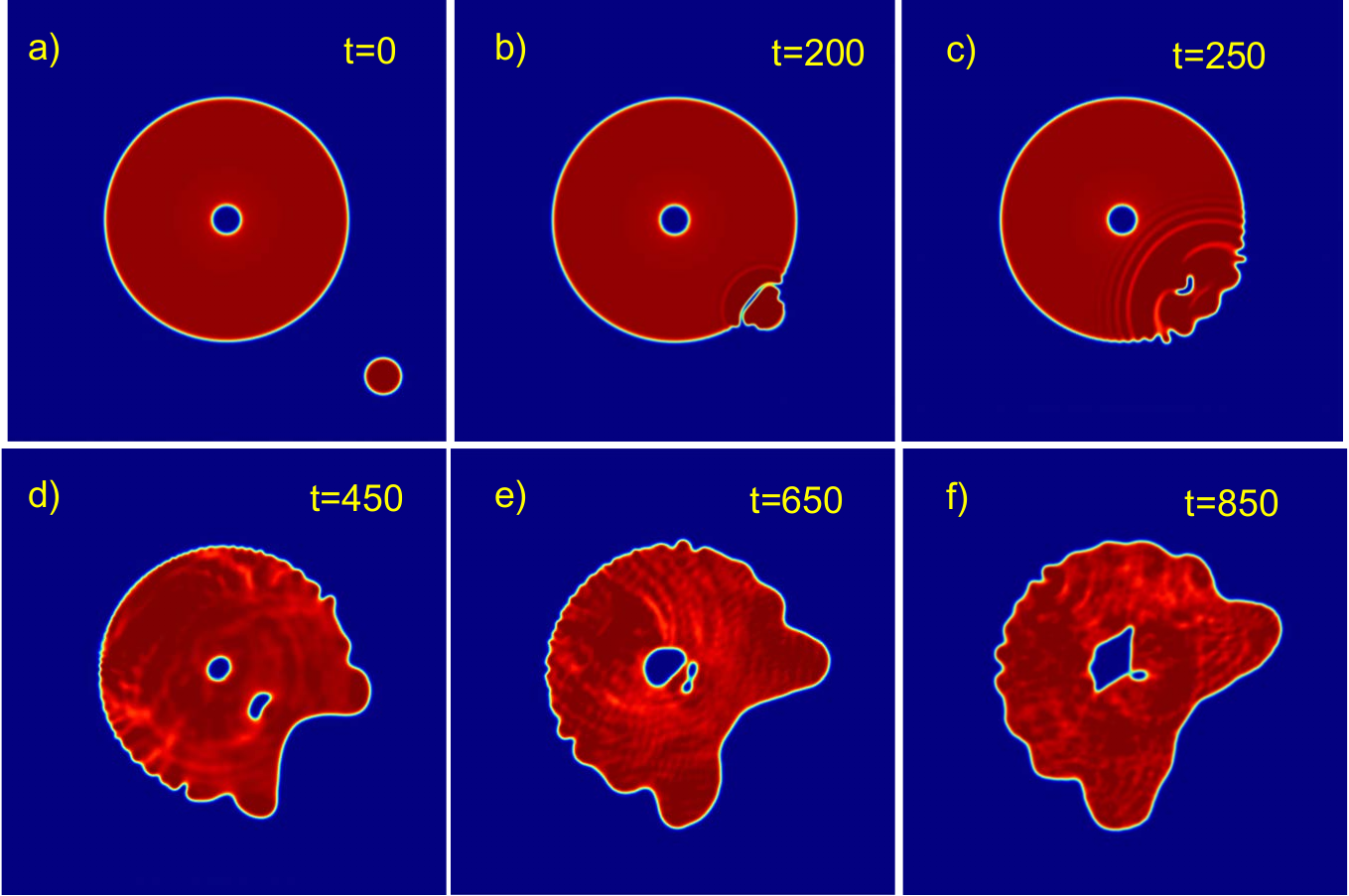}
\end{center}
\caption{A soliton-vortex collision generates a vortex-antivortex pair that moves toward the center and eventually disappears when
it reaches the central dark spot.
 All panels show a colormap of $|\Psi(x,y)|^2$ at different times, using the same conventions as in Fig.~\ref{fig1}.}
\label{fig11}
\end{figure}

The final example, shown in Fig.~\ref{fig11}, involves a larger impinging soliton, which enables the formation of a vortex-antivortex pair upon collision.
Initially, the vortex and antivortex phase singularities are located within the same density dip, but as the system evolves, the dips separate and become clearly 
distinguishable --- see Fig.~\ref{fig11} - e).
Eventually, they merge with the central vortex core, and the vortex-antivortex pair disappears from the flat-top region.
This simulation also demonstrates the robustness of flat-top vortex solitons: even after this rather violent interaction, the 
structure, characterized by a central density valley with $\ell = 2$ vorticity surrounded by a nearly flat-top region, remains intact.

\section{Conclusions and outlook}
\label{sec:conclusion}

We have presented an exploratory yet systematic analysis of form-preserving travelling wave solutions within a finite, symmetric, and effectively two-dimensional 
BEC forming a QD. Travelling waves in the form of rarefaction pulses and vortex-antivortex pairs appear as eigenstates of Eq.~(\ref{eqadim}) 
when boundary conditions correspond to the infinite system limit~\cite{paredes2025traveling}. In the case of flat-top finite solitons, these waves remain 
approximately form-preserving while they propagate far from the droplet edge, but they give rise to non-trivial phenomena upon interacting with the boundary. 

In Section~\ref{sec:bdy}, we showed that the typical outcome when a rarefaction pulse reaches the droplet edge is the emission of a bright soliton. For 
vortex-antivortex pairs, deflection near the boundary occurs due to boundary effects for the condensate phase,
 which can be modeled via an image antivortex-vortex pair. Depending 
on the initial separation between the phase singularities, different scenarios may arise: a flat-top soliton may be emitted, the singularities may reach the 
boundary and be expelled, or they may reverse direction and propagate backward as boundary excitations. All outcomes are accompanied by the excitation
of the hosting soliton.

In Section~\ref{sec:vortex}, we demonstrated that similar dynamics occur when dark excitations propagate within a vortex soliton, although the rotating 
background fluid alters their trajectories. Finally, in Section~\ref{sec:collisions}, we showed that both rarefaction pulses and vortex-antivortex pairs can 
be excited as a result of collisions involving bright solitons, either with or without vorticity. Altogether, our numerical simulations offer a broad, though 
not exhaustive, perspective on the qualitative behavior of travelling waves in the model considered.

Several directions remain open for future exploration. First, a more detailed analysis of the interactions between travelling waves --- either among themselves, 
with phase singularities, or with external inhomogeneities --- would be of interest. For instance, studies akin to 
those in~\cite{smirnov2015scattering,feijoo2017dynamics} could be carried out in the QD context, or the interaction of travelling waves 
with linear lattices could be investigated~\cite{chen2025dark}. 

Second, a natural question is how these dynamics generalize in more complex QD configurations, such as mixtures without component symmetry~\cite{kartashov2025double,pelayo2025droplet}, or in three-dimensional settings without confinement along any 
direction~\cite{kartashov2018three}. Finally, although experimentally challenging, it would be valuable to observe in the laboratory the different types of 
form-preserving travelling excitations discussed here. These could potentially be generated through phase imprinting techniques or via controlled 
soliton-soliton collisions. Quantum droplets have been proposed as a promising platform for the generation and characterization of 
self-trapped vortices~\cite{li2025can}. The creation and study of rarefaction pulses and vortex-antivortex pairs would be a 
natural and exciting subsequent step.

\section*{ACKNOWLEDGEMENTS}
This publication is part of the R\&D\&i project  PID2023-146884NB-I00,
funded by MCIN/AEI/10.13039/501100011033/. This work was also supported by 
grant ED431B 2024/42 (Conseller\i\i a de Cultura, Educaci\'on, Formaci\'on Profesional y
Universidades, Xunta de Galicia).
                                                        
\section*{DATA AVAILABILITY}

The data that support the findings of this article are openly
available \cite{data_availability}.

\bibliographystyle{apsrev}

\end{document}